\documentclass[12pt]{article}
\usepackage{amssymb}
\usepackage{bm}

\begin{document}
\title{Spiral symmetry and general Bloch's theorem}
\author{Z. C. Tu\\
Institute of Theoretical Physics\\
 The Chinese Academy of Sciences\\
 P.O.Box 2735 Beijing 100080, China}
\maketitle
\begin{section}{Abstract}
In this paper, spiral symmetry in cylindrical coordinate and
general Bloch's theorem induced from it are discussed. This
general Bloch's theorem is useful for considering the properties
related to single-walled carbon nanotubes.
\end{section}

\begin{section}{Translation symmetry in $\mathbb{R}^3$ and Bloch's theorem}
Above all, we go over traditional translation symmetry and Bloch's
theorem [1].\\
\\
Lattice vectors:
\begin{equation}{\bf R}_j=n_{j1}{\bf a}_1+n_{j2}{\bf a}_2+n_{j3}{\bf a}_3.\end{equation}
Hamiltonian:
\begin{equation}H=-\frac{\hbar^2}{2\mu} \nabla^2+V({\bf r});\quad  V({\bf r}+{\bf R}_j)=V({\bf r}).\end{equation}
Define translation operators $\mathcal{J}({\bf R}_j)\quad
(j\in\mathbb{N})$, which act on a function $f({\bf r})$ as:
\begin{equation}\mathcal{J}({\bf R}_j)f({\bf r})=f({\bf r}+{\bf
R}_j).\end{equation} We can easily proof:
\begin{equation}\label{b4}\mathcal{J}({\bf R}_j)\mathcal{J}({\bf R}_l)=\mathcal{J}({\bf
R}_l+{\bf R}_j)=\mathcal{J}({\bf R}_j+{\bf R}_l)=\mathcal{J}({\bf
R}_l)\mathcal{J}({\bf R}_j).\end{equation} Otherwise,
\begin{eqnarray}\mathcal{J}({\bf R}_j)Hf({\bf r})=-\frac{\hbar^2}{2\mu}
\mathcal{J}({\bf R}_j)\nabla^2f({\bf r})+\mathcal{J}({\bf
R}_j)V({\bf r})f({\bf r})\\
\nonumber =-\frac{\hbar^2}{2\mu}\nabla^2 \mathcal{J}({\bf
R}_j)f({\bf r})+V({\bf r}+{\bf R}_j)f({\bf r}+{\bf R}_j)\\
\nonumber =[-\frac{\hbar^2}{2\mu}\nabla^2+V({\bf r})]f({\bf
r}+{\bf R}_j)=H\mathcal{J}({\bf R}_j)f({\bf r}).\end{eqnarray}
Therefore $\{\mathcal{J}({\bf R}_j),H\}$ is the set of conserved
quantities. In this case, an eigenfunction of the Hamiltonian must
be an eigenfunction of the translation
operators.\begin{equation}H\psi({\bf r})=E\psi({\bf r})\Rightarrow
\mathcal{J}({\bf R}_j)\psi({\bf r})=\psi({\bf r}+{\bf
R}_j)=\lambda({\bf R}_j)\psi({\bf r}).\end{equation} On the one
hand, from Eq.(\ref{b4}), we know
\begin{equation}\label{b7}\lambda({\bf R}_j)\lambda({\bf R}_l)=\lambda({\bf R}_j+{\bf R}_l).\end{equation}
On the other hand, the electron density must be periodic, i.e.,
$|\psi({\bf r}+{\bf R}_j)|^2=|\psi({\bf r})|^2$, from which we
know \begin{equation}\label{b8}|\lambda({\bf
R}_j)|^2=1.\end{equation} Eqs(\ref{b7}) and (\ref{b8}) give
\begin{equation}\label{b9}\lambda({\bf
R}_j)=e^{i{\bm \kappa}\cdot {\bf R}_j}.\end{equation} Thus we have
Bloch's theorem:
\begin{equation}\label{b10}\psi({\bf r}+{\bf
R}_j)=e^{i{\bm \kappa}\cdot {\bf R}_j}\psi({\bf r}).\end{equation}
\end{section}
\begin{section}{Spiral symmetry in cylindrical coordinate and Bloch's theorem}
Hamiltonian in cylindrical coordinate:
\begin{equation}H=-\frac{\hbar^2}{2\mu} (\frac{\partial^2}{\partial\rho^2}+\frac{\partial}{\rho\partial\rho}+\frac{\partial^2}{\rho^2\partial\theta^2}+\frac{\partial^2}{\partial z^2})+V(\rho,\theta,z),\end{equation}
\begin{equation}\label{spiral}V(\rho,\theta,z)=V(\rho,\theta+j_1\vartheta,z+j_1\tau+j_2T),\end{equation}
where $\vartheta=2\pi/N, N\tau=MT$, $N,M\in\mathbb{N}$,
$j_1,j_2\in \mathbb{Z}$. Eq.(\ref{spiral}) is called the spiral
symmetry. Define vectors ${\bf r}=(\theta, z)$ and ${\bf
R}_j=(j_1\vartheta,j_1\tau+j_2T)$ in the space $[0,2\pi)\times
\mathbb{R}$. Define operators $\mathcal{J}({\bf R}_j)\quad
(j\in\mathbb{N})$, which act on a function $f(\rho;{\bf r})$ as:
\begin{equation}\mathcal{J}({\bf R}_j)f(\rho;{\bf r})=f(\rho;{\bf r}+{\bf
R}_j).\end{equation}
 We can easily proof:
\begin{equation}\label{bb1}\mathcal{J}({\bf R}_j)\mathcal{J}({\bf R}_l)=\mathcal{J}({\bf
R}_l+{\bf R}_j)=\mathcal{J}({\bf R}_j+{\bf R}_l)=\mathcal{J}({\bf
R}_l)\mathcal{J}({\bf R}_j),\end{equation} and
\begin{equation}\label{bb2}\mathcal{J}({\bf R}_j)V(\rho;{\bf r})=V(\rho;{\bf r})\mathcal{J}({\bf
R}_j).\end{equation} Otherwise,\begin{equation}\mathcal{J}({\bf
R}_j)(\frac{\partial^2}{\rho^2\partial\theta^2}+\frac{\partial^2}{\partial
z^2})f(\rho;{\bf
r})=(\frac{\partial^2}{\rho^2\partial\theta^2}+\frac{\partial^2}{\partial
z^2})\mathcal{J}({\bf R}_j)f(\rho;{\bf r}).\end{equation} Thus
\begin{equation}\label{bb3}\mathcal{J}({\bf R}_j)H=H\mathcal{J}({\bf
R}_j).\end{equation} We can obtain general Bloch's theorem
analogizing above section.
\begin{equation}H\psi(\rho;{\bf r})=E\psi(\rho;{\bf r})\Rightarrow
\mathcal{J}({\bf R}_j)\psi(\rho;{\bf r})=\psi(\rho;{\bf r}+{\bf
R}_j)=e^{i{\bm \kappa}\cdot {\bf R}_j}\psi(\rho;{\bf
r}).\end{equation} We set ${\bm \alpha}_1=(\vartheta, \tau)$,
${\bm \alpha}_2=(0,T)$, ${\bm \beta}_1=(N,0)$, ${\bm
\beta}_2=(-\tau N/T,2\pi/T)$, ${\bf G_j}=j_1{\bm \beta}_1+j_2{\bm
\beta}_2=(j_1N-j_2N\tau/T, 2\pi j_2/T)$, then we can obtain
\begin{equation}\label{plane}\psi(\rho;{\bf
r})=\sum\limits_{{\bf j}n}\phi_{n{\bf j}}(\rho)e^{i({\bm
\kappa}+{\bf G_j})\cdot {\bf r}}\end{equation}
\end{section}
\begin{section}{Discussion and Potential application}
Beginning with Eq.(\ref{plane}), we can continue to construct a
method similar to plane wave expansions [2]. We believe general
Bloch's theorem will be useful for consider the properties of
single-walled carbon nanotubes for their spiral symmetry [3][4].
\end{section}
\begin{section}{Reference}
Email: tzc@itp.ac.cn

[1] J. Callaway, {\it Quantum Theory of the Solid State} (Academic
Press, Inc., London, 1991).

 [2]Z. C. Tu, unpublished.

 [3] R.
Saito, M. S. Dresselhaus, and G. Dresselhaus, {\it Physical
Properties of Carbon Nanotubes} (Imperial College Press, London,
1998).

[4]Z. C. Tu and Z. C. Ou-Yang, cond-mat/0211658.

\end{section}
\end{document}